\documentclass{jkas}

\def\year{2015} 
\def\month{April} 
\def\volume{00} 
\def\issue{0} 
\def\beginpage{1} 
\def\endpage{99} 
\def\received{---} 
\def\accepted{---} 

\date{Received \received ; accepted \accepted}

\title{
{\it Euclid} Asteroseismology and Kuiper Belt Objects
}

\author[1]{A.~Gould}
\author[2,3,4]{D.~Huber}
\author[2,4]{D.~Stello}

\affil[1]{Department of Astronomy Ohio State University,
140 W.\ 18th Ave., Columbus, OH 43210, USA 
\email{gould@astronomy.ohio-state.edu }}
\affil[2]{School of Physics, University of Sydney, NSW 2006, Australia; 
\email{dhuber,stello@physics.usyd.edu.au}}
\affil[3]{SETI Institute, 189 Bernardo Avenue, Mountain View, CA 94043, USA}
\affil[4]{Stellar Astrophysics Centre, Department of Physics and
Astronomy, Aarhus University, Ny Munkegade 120, DK-8000 Aarhus C, Denmark}

\def\corrauthor{
Andrew Gould
}

\def\runningauthor{
Andrew Gould, Daniel Huber, Matthew Penny, Dennis Stello
}

\def\runningtitle{
Euclid Asteroseismology and KBOs
}

\def\keywords{
astrometry -- gravitational microlensing  -- stars: oscillations -- Kuiper belt
}

\def\abstracttext{
{\it Euclid}, which is primarily a dark-energy/cosmology mission, may
have a microlensing component, consisting of perhaps four dedicated
one-month campaigns aimed at the Galactic bulge.  We show that such
a program would yield excellent auxilliary science, including
asteroseimology detections for about 100,000 giant stars,
and detection of about 1000 Kuiper Belt Objects (KBOs), down to 
2--2.5 mag below the observed break in the KBO luminosity function at 
$I\sim 26$.  For the 400 KBOs below the break, {\it Euclid} will measure
accurate orbits, with fractional period errors $\lesssim 2.5\%$.
}

\def\au{{\rm AU}}

\def\vega{{\rm vega}}

\def\kms{{\rm km}\,{\rm s}^{-1}}

\def\muas{{\mu\rm as}}

\def\max{{\rm max}}

\def\ast{{\rm ast}}
\def\eff{{\rm eff}}

\def\rd{{\rm read}}
\def\single{{\rm single}}
\def\phot{{\rm phot}}
\def\unsat{{\rm unsat}}

\def\pix{{\rm pixel}}

\def\apj{{ApJ}}

\def\apjs{{ApJS}}
\def\aap{{A\&A}}

\def\mnras{{MNRAS}}

\begin{document}
\jkashead 


\section{{Introduction}
\label{sec:intro}}

In two earlier papers, we pointed out that {\it WFIRST} microlensing
observations toward the Galactic bulge would automatically yield
a treasure trove of asteroseismic \citep{wfirstastro} and Kuiper Belt Object
(KBO) \citep{wfirstkbo} data.  These papers contained detailed analytic
calculations that permit relatively easy scaling to other missions
and experiments.  One very relevant mission is {\it Euclid},
which is presently scheduled to be launched in 2020.  Unlike 
{\it WFIRST}, {\it Euclid} does not yet have a microlensing component,
but such a component is being actively discussed.  

Based on a most naive assessment, {\it Euclid} would appear 
to be much less effective in extracting non-microlensing science
from microlensing data than {\it WFIRST}.  {\it Euclid} has
1/2 the telescope diameter of {\it WFIRST}, 2.7 times larger linear
pixel scale, 1/4 as many visits to each target, and 1/2 the
angular area of the survey.

However, as we show, such naive assessment would be quite wrong since
{\it Euclid} will be able to detect oscillations in giants that are
about 0.8 mag brighter than for
{\it WFIRST}, i.e., to about 0.8 mag above the red
clump.  Hence, it will obtain asteroseismic measurements
for roughly 100,000 stars.

For KBOs, {\it Euclid} benefits greatly by having an optical channel in
addition to its primary infrared (IR) channel, even though it is
expected that the optical exposure time will be 3 times smaller than
for the IR.  The optical observations gain substantially from their
smaller point spread function (PSF) as well as the fact that KBOs
(unlike stars) do not suffer extinction.  Hence, {\it Euclid} will also
be a powerful probe of KBOs.

\section{{Euclid Characteristics}
\label{sec:char}}

We adopt {\it Euclid} survey characteristics from \citet{penny13},
with some slight (and specified) variations.  In each case, for
easy reference, we place the corresponding assumed {\it WFIRST}
parameters in parentheses.  Mirror diameter $D=1.2\,$m ($2.4\,$m);
pixel size $p=0.3^{\prime\prime}$ ($0.11^{\prime\prime}$), detector size
8k$\times$8k (16k$\times$16k), median wavelength $\lambda=1.7\,\mu$m
($1.5\,\mu$m), photometric zero point $H_\vega=23.5$ (26.1),
exposure time 52s (52s),  effective background (including read noise,
$H_{\vega,\rm sky}=20.0\,\rm arcsec^{-2}$, and dark current) $209\,e^-$ ($341\,e^-$),
full pixel well $n_\max = 10^5$ ($10^5$),
and single read time $t_\single = 2.6\,$s ($t_\single = 2.6\,$s).
In fact, \citet{penny13} do not specify a read time, so we use
the same value as for {\it WFIRST} to simplify the comparison.
Also, \citet{penny13} adopt an exposure time of 54 s, but we use
52 to again simplify the comparison.  Finally, \citet{penny13}
list $n_\max=2^{16}$, but this appears to be in error.  In any
case, since the detector is very similar to {\it WFIRST}, these
two numbers should be the same.

In our calculations we first consider simple 52 s exposures, but later
take account of the fact that five such exposures will be carried
out in sequence over $\sim 285\,$s.

\section{{Asteroseismology}
\label{sec:aster}}
\subsection{{Bright-star photometry}
\label{sec:phot}}
 From Equation (16) of \citet{wfirstastro}, the fractional error
(statistical) in the log flux $F$ (essentially, magnitude error) is
\begin{equation}
\sigma (\ln F) = [\pi n_\max(3-Q)]^{-1/2}r_\unsat^{-1};
\quad r_\unsat \equiv {\theta_\unsat\over p}
\label{eqn:sntotphot}
\end{equation}
where $p$ is the pixel size, $\theta_\unsat$ is the radius of the
closest unsaturated pixel (in the full read), $n_\max$ is the
full well of the pixel, 
\begin{equation}
Q = N_\rd^{-1} + 3(2 N_\rd)^{-2/3},
\label{eqn:qdef}
\end{equation}
and $N_\rd$ is the number of non-destructive reads.  

When comparing {\it WFIRST} and {\it Euclid}, the most important factor is $r_\unsat$.
To determine how this scales with mirror size, pixel size, exposure
time $t$, throughput $f$, and mean wavelength $\lambda$,
we adopt a common PSF function of angle $\theta$, $A(D\theta/\lambda)$,
which is scaled such
that $\int dx 2\pi x A(x)\equiv 1$.  
Then, if a total of $K$ photons fall on
the telescope aperture, the number falling in a pixel centered at $\theta$ 
is
\begin{equation}
n_\pix = p^2 Kf A(D\theta/\lambda) (D/\lambda)^2
\label{eqn:npixel}
\end{equation}
Setting $n_\pix = n_\max$, we derive
\begin{equation}
r_\unsat = {A^{-1}[n_\max\lambda^2/(Kf D^2 p^2)]\over pD/\lambda}.
\label{eqn:runsat}
\end{equation}
Then noting that $K = k f N_\rd t_\single D^2$, where
$k$ is a constant, we obtain
\begin{equation}
r_\unsat = {A^{-1}[n_\max\lambda^2/(k f t_\single  N_\rd D^4p^2)]\over pD/\lambda}
\label{eqn:runsat2}
\end{equation}
Finally, noting that in the relevant range, a broad-band Airy profile
scales $A(x)\propto x^{-3}$, we find a ratio of {\it Euclid} (E) to 
{\it WFIRST} (W) unsaturated radii,
\begin{equation}
{r_{\unsat,E} \over r_{\unsat,W}} =
\biggl[{(N_\rd D\lambda f/p)_E \over (N_\rd D\lambda f/p)_W}\biggr]^{1/3}
\simeq 0.42 .
\label{eqn:runsatrat}
\end{equation}
In making this evaluation we note that 
$(D_E/D_W)=0.50$,
$(p_E/p_W)=2.73$,
$(\lambda_E/\lambda_W)=1.13$, and
$(f_E/f_W) = 10^{0.4(23.52-26.1)}/(D_E/D_W)^2=0.365$.  The last number
may be somewhat surprising.  It derives mainly from {\it WFIRST}'s 
broader passband ($1\,\mu$m vs.\ $0.6\,\mu$m and its gold-plated
(so infrared optimized) mirror.
The other ratio of factors is
\begin{equation}
{3 - Q_E\over 3 - Q_W} = 1 .
\label{eqn:qeval}
\end{equation}
Therefore, we conclude that in the regime that the central pixel
is saturated in a single read, a sequence of 5 {\it Euclid} exposures 
(requiring about 285 s including readout) yields a factor 1.06 increase
in photometric errors compared to a single {\it WFIRST} exposure (lasting 52 s).

However, whereas this regime applies to stars $H<11.6$ for {\it WFIRST} 
($x$-intercept of lower curve of middle panel in Figure~1 of 
Gould et al.\ 2015), this break occurs at a brighter value for {\it Euclid}.  
To evaluate this offset, we rewrite Equation~(\ref{eqn:npixel}):
$N_\pix = kf N_\rd t_\single (D^2p/\lambda)^2 A(rDp/\lambda)$.
Hence, the offset (at fixed $r$) between the break points on 
the {\it WFIRST} and {\it Euclid} diagrams is
\begin{equation}
\Delta H = -2.5\log{[f D \lambda/p]_E/[f D \lambda/p]_H} 
= 2.8
\label{eqn:offset}
\end{equation}

That is, this boundary occurs at $H=11.6-\Delta H=8.8$, which is
significantly brighter than the majority of potential asteroseismic
targets.  However, inspection of that Figure shows that the same
scaling ($\sigma \propto F_H^{-1/3}$) applies on both sides of the
$H=11.6$ ``boundary''.  The reason for this is quite simple.  As we
consider fainter source stars $F_H$, the saturated region of course
continues to decline.  Since the Airy profile at large radii scales as
$A(x)\propto x^{-3}$, the radius of this region scales as
$r_\unsat\propto F_H^{1/3}$.  Hence, the number of semi-saturated
pixels (each contributing $n_\max$ to the total photon counts) scales
$\propto F_H^{2/3}$, which implies that the fractional error scales
$\propto F_H^{-1/3}$.  This breaks down only at (or actually, close
to) the point that the central pixel is unsaturated in a full read (at
which point the error assumes standard $\propto F_H^{-1/2}$ scaling).
That is, in the case of {\it Euclid}, the $F_H^{-1/3}$ scaling applies
to about $H<14.8-\Delta H = 12.0$, which is still toward the bright
end of potential targets.

\subsection{{Bright-star astrometry}
\label{sec:ast}}

 From \citet{wfirstastro}
\begin{equation}
\sigma(\theta) = {p\over\sqrt{3\pi n_\max \ln(1.78 N_\rd + 0.9)}},
\label{eqn:sigmatheta}
\end{equation}
in the saturated regime.
Hence, using the same parameters (for a single {\it Euclid} sub-exposure)
\begin{equation}
{\sigma(\theta)_E \over \sigma(\theta)_W} = 2.73.
\label{eqn:sigmathetarat}
\end{equation}
Then taking account of the fact that {\it Euclid} has five such sub-exposures,
we obtain
$\sigma(\theta)_E/\sigma(\theta)_W=1.22$, i.e., {\it Euclid} is very similar to
{\it WFIRST}.

However, as in the case of photometric errors, the boundary of
the regime to which this applies is $H<8.8$ for 
{\it Euclid} (compared to $H<11.6$ for {\it WFIRST}).  And, more importantly,
for astrometric errors, the functional form of errors does in fact
change beyond this break (see Figure 1 of Gould et al.\ 2015).

The reason for this change in form of astrometric errors can be understood
by essentially the same argument given for the form of
photometric errors in the previous section.  In the regime between
saturation of the central pixel in one read and $N_\rd$ reads, the
region $r<r_\unsat\propto F_H^{1/3}$ dominates the astrometric signal.
For astrometric signals, each pixel contributes to the (S/N)$^2$
as $n_\max/r^2$, and therefore the entire region contributes
\begin{equation}
\biggl({\rm S\over N}\biggr)^2\propto \int_1^{r_\unsat} {rdr\over r^2}
=\ln r_\unsat;
\qquad r_\unsat \propto F_H^{1/3}
\label{eqn:snast}
\end{equation}
Hence, between these two limits ($11.6<H<14.8$ for {\it WFIRST} and $8.8<H<12$
for {\it Euclid}) the astrometric errors should scale
$\sigma\propto(\ln F^{1/3})^{-1/2}\propto (\ln F)^{-1/2}$.  That is,
these errors should increase by a factor $\sqrt{\ln N_\rd}=1.73$, which
is indeed very similar to what is seen in Figure 1 of Gould et al.\ 2015.

\subsection{{Analytic Error Estimates for Bright Euclid Stars}
\label{sec:analytic}}

We summarize the results of Sections~\ref{sec:phot} and 
\ref{sec:ast} in the form of
analytic expressions for the photometric and astrometric errors for
bright stars as a function of magnitude for single-epoch {\it Euclid}
observations consisting of 5 52-second exposures.  The photometric
errors are,
\begin{eqnarray}
\sigma_\phot &=& 0.47\,{\rm mmag}\,10^{(2/15)(H-12)},\ (H<12)\nonumber\\
\sigma_\phot &=& 0.47\,{\rm mmag}\,10^{(1/5)(H-12)},\  (H>12)
\label{eqn:sigmaphotall}
\end{eqnarray}
The astrometric errors are
\begin{eqnarray}
\sigma_\ast &=& 73\,\muas\quad (4.4<H<8.8)\nonumber\\
\sigma_\ast &=& 73\,\muas\,
\sqrt{{\ln 10\over 2.5}(H-8.8) + 1}
\quad (8.8<H<12)\nonumber\\
\sigma_\ast &=& 145\,\muas
\,10^{(1/5)(H-12)}
\quad (H>12)
\label{eqn:sigmaastall}
\end{eqnarray}

\begin{table}[t!]
\begin{center}
\begin{tabular}{l c c c}        
\hline         
Parameter & KIC\,2437965 & KIC\,2425631 & KIC\,2836038 \\	
\hline		
$T_\eff$\ (K)     & 4356   & 4568 & 4775    \\
$\log g$\ (cgs)  & 1.765  & 2.207 & 2.460    \\
$\rm [Fe/H]$\ (dex)& 0.43 & $-0.10$ & 0.33     \\
$R (R_\odot)$     & 24.94  & 16.44 & 11.28    \\
$M_{H}$ (mag)    & $-3.14$ & $-2.38$ & $-1.60$    \\
$A_{H}/A_{Kp}$    & 0.45    & 0.45 & 0.45    \\
$\sigma_{E}$ (mmag)& 0.483 & 0.685 & 0.984   \\
Reference         &P14   & P14 & C14      \\
\hline		
\end{tabular} 
\caption{Fundamental properties and simulation parameters for
{\it Euclid} simulations.  P14 = \citet{apokasc}, C14 = \citet{casa14}.} 
\label{tab:params}
\end{center} 
\end{table}

\subsection{{Asteroseismic Simulations for Euclid}
\label{sec:sim}}

As for {\it WFIRST}, we evaluate the asteroseismic capabilities of {\it Euclid}
by its ability to recover the frequency of maximum
power $\nu_\max$ and the large frequency separation $\Delta\nu$. Both are key
quantities that can be used to estimate radii and masses for large
ensembles of giants, as demonstrated with the {\it Kepler} sample
(e.g., \citealt{kallinger10,hekker11,mosser12}).

\begin{figure}
\centering
\includegraphics[width=85mm]{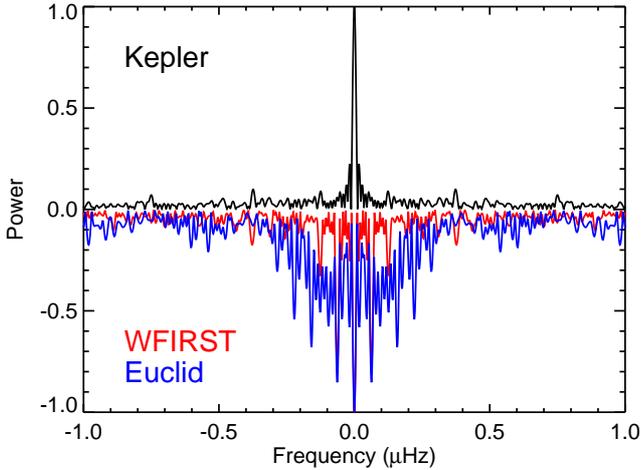}
\caption{Spectral window function for a typical {\it Kepler} time series 
(top panel, black) and after degrading the time series to a typical duty cycle 
expected for {\it Euclid} and {\it WFIRST} (bottom panel, blue, red).  
Note that while the {\it Euclid} peak is much broader than for {\it Kepler}
the power is still contained within
$\sim 0.5\,\mu$Hz, which is substantially narrower than
most spectral features of interest for most stars.}
\label{fig:swf}
\end{figure}

\begin{figure}
\begin{center}
\resizebox{85mm}{!}{\includegraphics{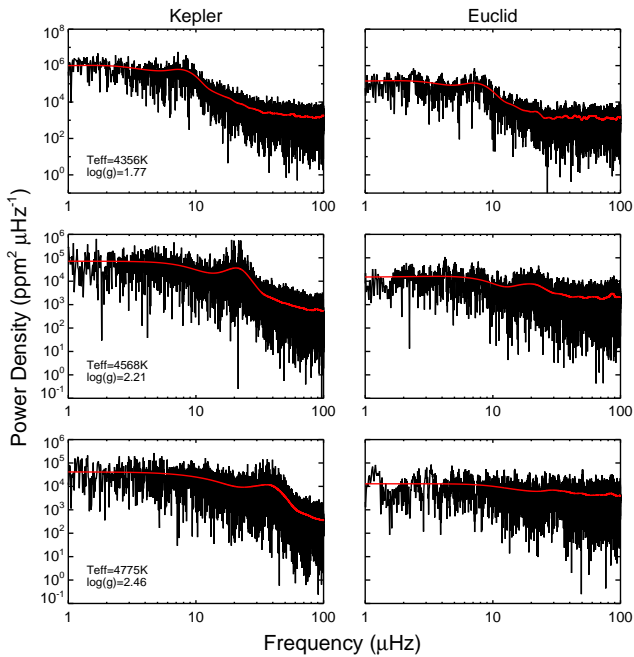}}
\caption{Power spectra of {\it Kepler} observations (left panels) and simulated 
{\it Euclid} observations (right panels) for three red giants in 
different evolutionary stages: high-luminosity red giant 
(top panels), 0.8 mag above red clump (middle panels) and 
red clump star (bottom panels). 
Red lines show the power spectra smoothed 
with a Gaussian with a full-width half-maximum of $2\Delta\nu$. Estimated 
stellar properties are given in the left panels, with a more complete
description given in Table~\ref{tab:params}.}
\label{fig:ps}
\end{center}
\end{figure}

\begin{figure}
\begin{center}
\resizebox{85mm}{!}{\includegraphics{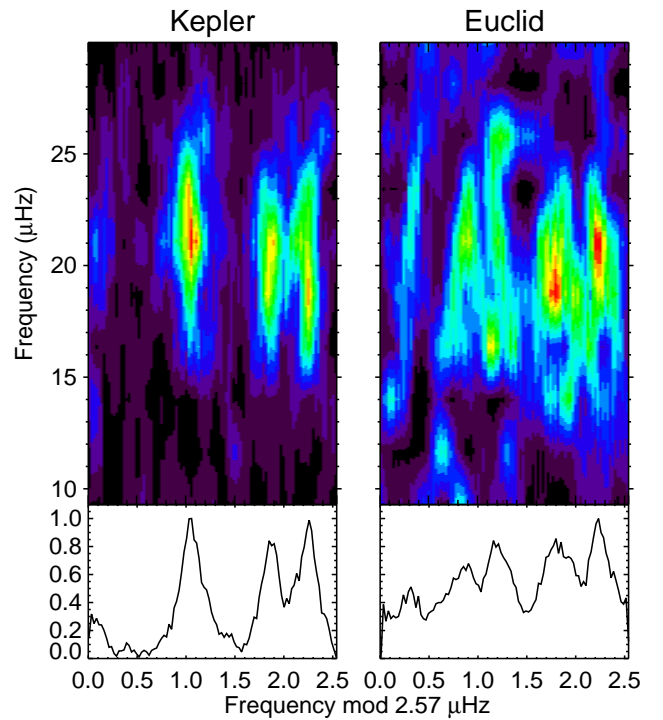}}
\caption{\'Echelle diagram for KIC\,2425631 (corresponding to
middle panel of Figure~\ref{fig:ps}) for {\it Kepler} (left)
and simulated {\it Euclid} (right) data. The abscissa is the
frequency offset relative to the beginning of each order, 
which are separated by (an adopted) large-frequency spacing
$\Delta\nu=2.57\,\mu$Hz.  Hence, for example,
at the frequency of maximum power $\nu_\max\sim 20\,\mu$Hz, the
ordinate corresponds to the eighth order.  In the {\it Kepler}
\'echelle, the three mode degrees $l=1,2,0$ from left to right are easily
discerned, which is not true of the {\it Euclid} \'echelle.  However,
the {\it Euclid} \'echelle does display clear vertical ``streaking'' which
is the signature that the adopted $\Delta\nu=2.57\,\mu$Hz
is the correct one.
}
\label{fig:echelle}
\end{center}
\end{figure}

Figures~\ref{fig:swf}, \ref{fig:ps}, are the {\it Euclid} analogs of
Figures 3, and 4 from \citet{wfirstastro} for {\it WFIRST},
while Figure~\ref{fig:echelle} illustrates the same physics as
Figure 5 from \citet{wfirstastro}.
In addition to incorporating
the analytic formulae summarized in Section~\ref{sec:analytic},
they also take account of the following assumptions about the
{\it Euclid} microlensing observations.  First, they assume an 18 minute
observing cycle (compared to 15 minutes for {\it WFIRST}).  Second, they
assume four 30-day observing campaigns (compared to six 72-day campaigns
for {\it WFIRST}).  Finally, we adopt a specific ``on-off'' (bold, normal)
schedule of ({\bf 30},124,{\bf 30},335,{\bf 30},181,{\bf 30}) days.
This schedule has been chosen to be consistent with the {\it Euclid}
sun-exclusion angle and to have two campaigns in each of the spring
and autumn (useful for parallaxes) but is otherwise arbitrary.

The area under the spectral window function in Figure~\ref{fig:swf}
is about 2.4 times larger than for {\it WFIRST} due to the fact that
the campaigns are a factor $72/30=2.4$ times shorter.  Nevertheless,
the FWHM of this envelope is still only about 500 nHz, which is
far less than the $\nu_\max\sim 8\,\mu$Hz for the brightest star
shown in Figure~\ref{fig:ps}.  Hence, it is only for extremely
bright stars that the width of this envelope will degrade the
measurement of $\nu_\max$.

Figure~\ref{fig:ps} is qualitatively similar to Figure 4 from 
\citet{wfirstastro}.  Note that as for the {\it WFIRST} simulations,
the {\it Kepler} time series has been shorted to the assumed
full duration (with gaps, 760 days) of the {\it Euclid} run to
allow a direct comparison of the effects of sampling.
As shown in Sections~\ref{sec:phot} and 
\ref{sec:analytic},
the individual photometric measurements have very similar
precision for {\it WFIRST} and {\it Euclid} in the relevant magnitude range.
However, {\it Euclid} has a factor $(18/15)\times (72/30) \times (6/4)=4.0$
times fewer of them, which leads to a factor 2.0 less sensitivity.
This means that the ``noise floor'' kicks in at a power density of
about $10^{3.3}\,\rm ppm^2\,\mu Hz^{-1}$ rather than $10^{3.0}$.  This
is the reason that in the bottom panel, the $\nu_\max$ peak is
not visible above the floor, whereas for {\it WFIRST} it is robustly
visible (middle panel of Figure~4 of \citealt{wfirstastro}).  
On the other hand, the $\nu_\max$ peak in the top
panel is about equally robust in simulated {\it WFIRST} and {\it Euclid} data.
The middle panel represents the approximate limit of {\it Euclid}'s ability
to measure $\nu_\max$.

Figure~\ref{fig:echelle} is an \'echelle diagram for the spectrum
shown in the middle panel of Figure~\ref{fig:ps}, i.e., the one just
described that is at the limit of {\it Euclid}'s ability to measure $\nu_\max$.
Figure~\ref{fig:echelle} shows that this star is also just above the
limit of {\it Euclid}'s ability to measure the other key asteroseismic
parameter, the large frequency spacing $\Delta\nu$.
In these diagrams, the abscissa designates frequency modulo
the adopted $\Delta\nu$,
while the ordinate is frequency.  Hence, they represent the power
spectrum divided into $\Delta\nu$-wide bins and stacked one above the
other.
If the adopted $\Delta\nu$ is correct (and the data are of sufficient
quality) then the diagram should look like a series of vertical streaks,
one for each mode degree $(l=1,2,0)$ comprising a series of overtones.
This is clearly the case for the original {\it Kepler} data.  
While the ridges
in the simulated Euclid data are heavily smeared out due to the window
function, an identification of the correct $\Delta\nu$ is still possible. Hence
both $\nu_\max$ and $\Delta\nu$ 
are measurable for this star. These quantities should
also be measurable for brighter giants because these have both larger
amplitude oscillations and smaller photometric errors, although for the most
luminous giants the frequency resolution for a typical 
{\it Euclid} observing run will be too low to resolve $\Delta\nu$.

\subsection{{Role of Euclid Parallaxes}
\label{sec:par}}

In \citet{wfirstastro}, we argued that {\it WFIRST} parallaxes could help
resolve ambiguities in the measurement of $\Delta\nu$
due to aliasing.  That is, from color-surface brightness relations,
one approximately knows the angular radius, which combined with
the measured parallax, gives the physical radius $R$.  By combining
$R$, $\nu_\max$, and general scaling relations, one can approximately
predict $ \Delta\nu$, and so determine which of the
alias-peaks should be centroided to find a more precise value.

However, this does not work for {\it Euclid}.  From Equation~(\ref{eqn:sigmaastall})
and the fact that there are a total of $\sim 10,000$ observations,
it follows that the parallax errors are
\begin{eqnarray}
\sigma(\pi) &=& 8\,\muas\,
\sqrt{{\ln 10\over 2.5}(H-8.8) + 1}
\quad (8.8<H<12)\nonumber\\
\sigma(\pi) &=& 15\,\muas
\,10^{(1/5)(H-12)}
\quad (H>12)
\label{eqn:sigmaastall2}
\end{eqnarray}
Hence, at the brightness of targets near the boundary of
measurability, $H\sim 13$, the parallax
errors are roughly 20\%, which provides essentially no information.
Even for the rarer targets at $H\sim 12$, the parallax measurements
do little more than confirm that the star is in the bulge,
which is already basically known in most cases.  Nevertheless, these
parallaxes are substantially better than Gaia parallaxes for the
same stars and so could be useful for other purposes.

\section{{Kuiper Belt Objects}
\label{sec:kbo}}

\citet{wfirstkbo} carried out analytic calculations to assess 
how well {\it WFIRST} could detect and characterize Kuiper Belt Objects (KBOs),
including orbits, binarity, and radii.  Here we apply these
analytic formulae to {\it Euclid}.  

In contrast to {\it WFIRST}, {\it Euclid} will
observe in two channels simultaneously, using an optical/IR
dichroic beam-splitter.  Because the microlensing targets will
be heavily extincted in the optical, and also because bandwidth
considerations restrict the optical downloads to once per hour
(vs.\ once per 18 minutes for the IR), the optical data will
provide only supplementary information for microlensing events.
Similarly, for the heavily reddened (and intrinsically red) asteroseismology
targets, the optical images will also be of secondary importance, and
so were not considered in Section~\ref{sec:aster}.  However, KBOs
lie in front of all the dust and hence both the optical and IR channels
should be considered.

\subsection{{Euclid IR Observations of KBOs}
\label{sec:IR}}

We remind the reader that, in contrast to asteroseismology
targets, KBOs are below the sky level 
and therefore are in a completely different
scaling regime.  In order to make use of the analytic
formulae of \citet{wfirstkbo}, we first
note that the {\it Euclid} point spread function (PSF) is slightly more undersampled
than that of {\it WFIRST} by a factor 
$(0.30/0.11)\times(1.2/2.4)\times(1.5/1.7)=1.2$.  We therefore adopt
an effective sky background
of $6B$, where $B=209\,e^-$ is the effective background
in a single pixel (compared to $3B$ for the oversampled limit).  That is,
$H_{\rm sky} = 23.5 - 2.5\log(6B/52)=20$, i.e., 
1.7 mag brighter than for {\it WFIRST}.
Then, following Equation (3) of \citet{wfirstkbo}, we estimate the astrometric
precision of each measurement as
\begin{equation}
\sigma_{\rm ast} = \sqrt{2}{\sigma_{\rm psf}\over {\rm SNR}} = 
{\rm 212\, mas\over SNR},
\label{eqn:kboast}
\end{equation}
i.e., exactly twice the value for {\it WFIRST} (because the mirror is two
times smaller).  Here, the signal-to-noise ratio (SNR) of 5 co-added consecutive
observations is given by
\begin{equation}
{\rm SNR} = 10^{0.4(H_{\rm zero} - H)}, \qquad H_{\rm zero} = 22.6.
\label{eqn:snr}
\end{equation}
Comparing this to Equation~(1) of \citet{wfirstkbo}, we see that
$H_{\rm zero}$ is 3.5 mag brighter than for {\it WFIRST}.

Before continuing, we remark on the issue of smeared images, which
is potentially more severe for {\it Euclid} than {\it WFIRST} because the
full duration of the 5 co-added exposures is about 5.5 times longer
than the {\it WFIRST} exposure.  However, due to the shorter duration of
the observing window and the fact that this window has one end
roughly at quadrature (and the other 30 days toward opposition),
the typical relative
velocities of the KBO and satellite are only about $5\,\kms$,
which corresponds to $\sim 10\,\muas\,{\rm min}^{-1}$, which is still
too small to substantially smear out the PSF, even in $\sim 5\,$min
exposures.

Although the area of the {\it Euclid} field is only about half the size
of the {\it WFIRST} field, the fact that the campaign duration is only
40\% as long together with the very slow mean relative motion of the
KBOs (previous paragraph) implies that this is even less of an issue
than for {\it WFIRST}, which \citet{wfirstkbo} showed was already quite
minor.  We therefore ignore it.

As with {\it WFIRST}, we assume that only 90\% of the observations are usable,
due to contamination of the others by bright stars.  This balances
the opposite impact of two effects: larger pixels increase the chance
that a given bright star lands on the central pixel, while brighter
$H_{\rm zero}$ decreases the pool of bright stars that can contribute
to contamination.  We also assume that 10\% of the observing time
is spent on other bands.  These will provide important color information
but are difficult or impossible to integrate into initial detection
algorithms.  Hence the total number of epochs per campaign is 
$N_{\rm cam}=1920$, i.e., a factor $\sim 3$ smaller than for {\it WFIRST}.

To determine the minimum total SNR required for a detection, we
must determine the number of trials.  We begin by rewriting
Equation (18) of \citet{wfirstkbo} as
\begin{equation}
N_{\rm try} = 7\times 10^{14}
\biggl({\Delta t\over {\rm day}}\biggr)^5
\biggl({N_{\rm pix}\over 3\times 10^9}\biggr)
\biggl({\theta_{\rm pix}\over 110\,{\rm mas}}\biggr)^{-4}
\label{eqn:try}
\end{equation}
where $N_{\rm pix}= 3\times 8000^2 = 2\times 10^8$ is the total
number of pixels in 3 {\it Euclid} fields, $\Delta t=30\,$day is the
duration of a {\it Euclid} campaign, and $\theta_{\rm pix}=300\,$mas
is the size of a {\it Euclid} pixel.  Therefore $N_{\rm try}=2\times 10^{19}$.
This is almost five orders of magnitude smaller than the corresponding
number for {\it WFIRST}.  Hence, the challenges posed by limitations
of computing power that were discussed by \citet{wfirstkbo}
are at most marginally relevant for {\it Euclid}.  Therefore, we ignore them
here.

Then to find the limiting SNR at which a KBO can be detected, we solve
Equation (17) from \citet{wfirstkbo}, i.e.,
\begin{equation}
{\rm SNR} \gtrsim N_{\rm cam}^{-1/2}\sqrt{2\ln{N_{\rm try}\over {\rm SNR}}
- \ln N_{\rm cam}} = 0.22,
\label{eqn:snr2}
\end{equation}
where $N_{\rm cam}$ is the of observations during a single-season
observing campaign.
Hence, the limiting magnitude is $H\lesssim 24.2$.

\subsection{{Euclid Optical Observations of KBOs}
\label{sec:optical}}

To compare optical with IR observations, we first note that the
field of view is the same, but with pixels that are smaller by a factor 3.
This means that the optical data are nearly critically sampled
(100 mas pixels and 180 mas FWHM).  The ``RIZ'' band is centered
in $I$ band and has a flux zero point of $I_\vega=25.1$.  As noted above,
band-width constraints limit the number of optical 
images that can be downloaded (in addition to the primary IR images) to
one per hour (for each of the three fields).  

Taking into account the read noise of $4.5\,e^-$, a sky background
of $I=21.0\,\rm arcsec^{-2}$, and the nearly critical
sampling, we find the analog of Equation~(\ref{eqn:snr}) to be
\begin{equation}
{\rm SNR} = 10^{0.4(I_{\rm zero} - I)}, \qquad I_{\rm zero} = 27.4
\label{eqn:snri}
\end{equation}
for 270 s exposures.  Even allowing for the fact that there are only
30\% as many $I$-band observations and that the typical color of
a KBO is $I-H\sim 1$, this still represents an  improvement of
$27.4-22.6-1+2.5\log(0.3) = 2.5$ magnitudes relative to $H$ band.
Thus, the $I$-band observations will be the primary source of
information about KBOs and we henceforth focus on these.

An important benchmark for understanding the relative sensitivity
of {\it WFIRST} and {\it Euclid} is that at the KBO luminosity function break
($R=26.5$, $I=26.0$, $H=25.1$), we have SNR$\sim 2.5$ for {\it WFIRST}
and SNR$\sim 3.6$ for {\it Euclid}.

For fixed field size and $\Delta t$, the number of trials 
(Eq.~(\ref{eqn:try})) scales $N_{\rm try}\propto \theta_{\rm pix}^{-6}$,
i.e., $N_{\rm try}\sim 10^{22}$, 
a factor $10^{2.9}$ higher for the optical than the IR.
This implies that reaching the theoretical detection
limit will require roughly $10^{26}$ floating point operations (FLOPs).
While this is three orders of magnitude lower than {\it WFIRST}, it is not
trivially achieved \citep{wfirstkbo}.  For the moment we evaluate
the detection limit assuming that it can be achieved and then
qualify this conclusion further below.  Assuming that 10\% of observations
are lost to bright stars and then
solving Equation~(\ref{eqn:snr2}) yields SNR$\gtrsim 0.39$, i.e.,
a detection limit of $I\lesssim 28.4$.  Considering that the
break in the KBO luminosity function is $R\sim 26.5$ and that typically
$R-I\sim 0.45$, this is about 1.5 mag below the break.  Hence, {\it Euclid}
optical observations will be a powerful probe of KBOs.

\subsection{{Detections}
\label{sec:detect}}

The main characteristics of the {\it Euclid} optical KBO survey can now
be evaluated by comparing to the results of \citet{wfirstkbo}.
The total number detected up to the break (and also the total number
detected per magnitude between the break and the faint cutoff) will
be $(4/6)/2.1 = 32\%$ smaller than for {\it WFIRST}
(Figure~4 from \citealt{wfirstkbo}) because there are 4 campaigns
(rather than 6) and the survey area is a factor 2.1 times smaller.

Thus, there will be a total of 400 KBOs discovered that are brighter
than the break and 530 per magnitude fainter than the break.
In fact, this distribution is only known to be flat for about 1.5 mag.
{\it Euclid} observations will reach about 2.4 mag below the break,
provided that the computational challenges can be solved.  However,
as we discuss in Section~\ref{sec:compute}, even if they cannot be solved, this
will pull back the magnitude limit by only a few tenths.

Hence, {\it Euclid} will discover at least 800 KBOs down to the point
that the KBO luminosity function is measured, and perhaps a few
hundred beyond that.

\subsection{{Orbital Precision}
\label{sec:orbit}}

To determine the precision of the orbit solutions, we first note
that the PSF is almost exactly the same size (both mirror and
observing wavelength are half as big).  Hence, Equation~(12) from
\citet{wfirstkbo} remains valid:
\begin{eqnarray}
& &\sigma(v_r)\over v_\oplus\sqrt{\Pi}\\
&=&{3.8\times 10^{-3}\over {\rm SNR}}
\biggl( {N \over 5600} \biggr)^{-1/2}
\biggl( {\Delta t \over 72 \rm d} \biggr)^{-3}
\biggl( {r \over 40\,\au} \biggr)^{5/2}\\
&=&{0.15\over {\rm SNR}}
\biggl( {N \over 650} \biggr)^{-1/2}
\biggl( {\Delta t \over 30 \rm d} \biggr)^{-3}
\biggl( {r \over 40\,\au} \biggr)^{5/2}.
\label{eqn:vr}
\end{eqnarray}
Here $v_r$ is the instantaneous KBO radial velocity,
$r$ is the KBO distance, $\Pi = \au/r$, $v_\oplus$ is Earth's
orbital velocity, and $N$ is the number of contributing images.
\citet{wfirstkbo} argued that because $v_r$ was by far the worst
measured KBO Cartesian coordinate, all orbital-parameter errors
would scale with this number, and in particular for the period $P$
(his Equation~(13)),
\begin{equation}
{\sigma(P)\over P} \simeq 
{3 v_r\over v_\oplus\Pi^{1/2}}
{\sigma(v_r)\over v_\oplus\Pi^{1/2}}.
\label{eqn:period}
\end{equation}
Hence, at fixed SNR (per observation), {\it Euclid} period errors are a factor
40 larger than for {\it WFIRST}.  The primary reason for this is the factor 2.4
shorter observing campaign, which enters as the third power.  In addition,
there are about 7 times fewer observations, which enters as the square root.
At the break ($I\sim 26$), we have from Equation~(\ref{eqn:snri}), 
SNR$\simeq 3.6$.  Following \citet{wfirstkbo}, we adopt 
$v_r\sim 0.2 v_\oplus\Pi^{1/2}$, and derive $\sigma(P)/P\sim 2.5\%$.
This means that orbital precisions for KBOs below the break will
be good enough to determine orbital families, but in most cases
not good enough to detect detailed subtle structures.

\subsection{{Binaries}
\label{sec:binaries}}

Microlensing-style observations can detect KBOs through two distinct
channels: resolved companions and unresolved companions detected
from center-of-light motion \citet{wfirstkbo}.

For resolved companions, the situation for {\it Euclid} is essentially
identical to {\it WFIRST}.  That is, the resolution is the same, and
gain in sensitivity from reduction in trials is nearly the
same.  Therefore, {\it Euclid} can detect binaries down to $I\lesssim 29.3$.
Note that because of the relatively small number of trials, this
limit is independent of whether the computational challenges to
reaching the $I=28.4$ detection limit can actually be achieved.

Once the companion is detected, its proper motion (relative to the
primary) can be measured with a precision (Eq.~(21) of \citealt{wfirstkbo})
\begin{equation}
\sigma(\Delta\mu) = \sqrt{24\over N}{\sigma_{\rm ast}\over \Delta t}
= {250\,\rm mas\, yr^{-1}\over SNR},
\label{eqn:sigmamu}
\end{equation}
where $\sigma_{\rm ast} = 105\,$mas is the Gaussian width of the PSF.
This is a factor $\sim 7$ larger error than for {\it WFIRST}.  Considering
that single-epoch SNR for {\it Euclid} is 3.6/2.5 = 1.44 better than for
{\it WFIRST}, Equation~(22) of \citet{wfirstkbo} becomes
$\Delta\mu \sim 83({\rm SNR})^{1/2}\eta^{-1/2}$ where $\eta$ represents
the binary separation relative to the Hills-sphere radius.  Therefore,
a 3-sigma detection of the proper motion requires 
$\eta\lesssim ({\rm SNR}/4.3)^3$.  Hence, for example, at the break,
$\eta\lesssim 0.6$.  Since for KBOs at the break (diameter $D\sim 50\,$km),
and for orbits with 
$a\sim 40\,\au$, the Hills sphere is at roughly $7^{\prime\prime}$,
this still leaves plenty of room for detections with proper motion
measurements.  Such measurement can be used to statistically constrain
the masses of the KBOs.  However, the main problem is that most
observed KBO companions are at much closer separations, indeed too
close to be resolved.

\citet{wfirstkbo} therefore investigated how well these could be
detected from center-of-light motion.  This requires that the
companion be separated by less than a pixel (otherwise not unresolved),
the period be shorter than the duration of observations (otherwise
center-of-light internal motion cannot be disentangled from center-of-mass
motion around the Sun), and high-enough ($\gtrsim 7\,$ sigma) detection
to distinguish from noise spikes.

 From Equation~(23) of \citet{wfirstkbo}
for such detection, we have
\begin{equation}
{\rm SNR}\gtrsim 7\sqrt{2\over 650}{p\over f_{\rm cl}\theta_c}
\gtrsim 4.85 {p\over\theta_c}\biggl({f_{\rm cl}\over 0.08}\biggr)^{-1},
\label{eqn:bindet}
\end{equation}
where $\theta_c$ is the binary projected separation and $f_{cl}$ is the
ratio of center-of-light to binary motion, which has a broad peak
$0.07\lesssim f_{\rm cl}\lesssim 0.09$ \citep{wfirstkbo}.  This corresponds
to $I<25.6$, i.e., almost a half mag above the break.

The requirement that the
period obey $P_c<\Delta t$ implies that
$\eta = (a/\au)^{-1}(P_c/{\rm yr})^{2/3} = 0.0047$.  Hence, imposing
$\theta_c=p$ (maximum unresolved orbit) yields a 
KBO diameter $D=pD_\odot/\eta = 145\,$km, corresponding to $I\sim 23.5$,
i.e., 2.5 mag above the break.

\begin{figure}
\centering
\includegraphics[width=90mm]{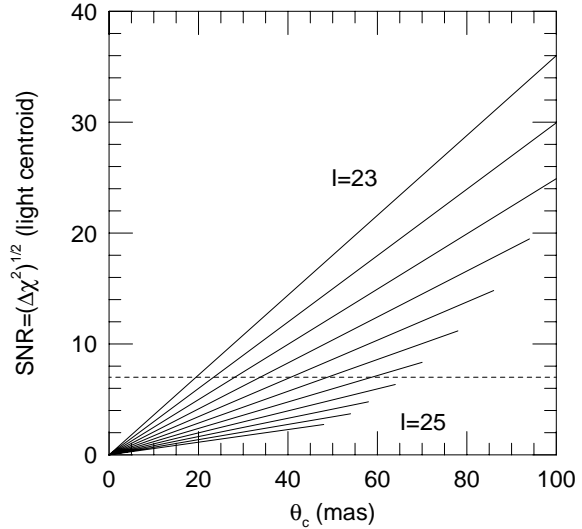}
\caption{Signal-to-noise ratio $[(\Delta\chi^2)^{1/2}]$ for orbiting
KBOs at a range of separations that are less than the {\it Euclid} optical pixel
size $\theta_c<p=100\,$mas.  The KBO brightness ranges from $I=23$
to $I=25$ as indicated.  Tracks end to the right when the orbital
period $P=\Delta t=30\,$days, the duration of an observing campaign.
Longer period orbits would have substantially lower signal.  The
calculations assume $f_{\rm cl}=8\%$ (see Fig.~1 of \citealt{wfirstkbo}).
The region of partial (broad) sensitivity $I<24$ ($I<23$)
will have only about 25 (6) detected KBOs, so not many binaries
will be detected via this channel (in strong contrast to {\it WFIRST}.
}
\label{fig:it}
\end{figure}

These analytic estimates imply that detection of binaries from light-centroid
motion is much more difficult than 
with {\it WFIRST}, a conclusion that is confirmed
by Figure~\ref{fig:it}, which is the analog of Figure~2 from \citet{wfirstkbo}.
It shows that such detections are impossible for $I>24$ and begin to
cover a wide range of separations only for $I<23$.  The total numbers
of expected KBO detections in these ranges are 
$\sim 25$ and $\sim 6$, respectively.  Hence, there will be very
few binaries detected through the light-centroid channel.

\subsection{{Occultations}
\label{sec:occult}}

Similarly, {\it Euclid} will detect very few KBO occultations compared
to the ${\cal O}(1000)$ occultations expected for {\it WFIRST}.
First, the total number of (epochs)$\times$(sky area) is smaller
by a factor 27.  Second, the exposure times are longer by a factor $\sim 5$.
Since the occultations are typically shorter than the {\it WFIRST} exposure
time, this means that the signal is $\sim 5$ times weaker for otherwise
equivalent stars.  This factor does also imply that the number
of occultations is in principle larger by this same factor $\sim 5$
for similar trajectories, but having more occultations is of no help
if they are each unobservably faint.
Finally, because the fields are moderately extincted, the occulted
stars are fainter in the $I$-band than in $H$.  Therefore, we conclude
that {\it Euclid} KBO occultations will not provide much information about
KBOs.

\subsection{{Computational Challenges}
\label{sec:compute}}

As mentioned in Section~\ref{sec:optical}, a total of $10^{26}$ FLOPs
would be required to reach the detection limit of $I=28.4$.  By contrast,
\citet{wfirstkbo} argued that it would be straightforward today
to carry out $10^{23.5}$ FLOPs per year, and that perhaps in 10 years
Moore's Law might raise this number to $10^{25}$.  This would leave
a shortfall of $q=10^{2.5}$ or $q=10^1$, in the two cases, respectively.
If these shortfalls could not be overcome, then according to the argument
given in Section~6 of \citet{wfirstkbo}, this would lead to a cutback
of the magnitude limit by $\Delta I\sim (2.5/12)\log q$, which is
0.5 or 0.2 mag in the two cases, respectively.  Note that in both
cases, the limit would still be beyond the current limit where the
luminosity function is measured (i.e., 1.5 mag beyond the break).
Hence, even if such computational challenges cannot be overcome,
this will not compromise {\it Euclid}'s capability to explore new regimes
of KBO parameter space.

\section{{Conclusion}
\label{sec:conclude}}

We have applied to the {\it Euclid} mission
analytic formulae that were previously derived for {\it WFIRST} 
in order to assess the implications of microlensing observations
for asteroseismology and KBO science.  In contrast to {\it WFIRST},
{\it Euclid} does not at present have a dedicated microlensing component.
We have therefore used the survey parameters presented by \citet{penny13}
as a guideline to a microlensing program that is being actively considered.

We find that for asteroseismology, {\it Euclid} observations are nearly as good
as {\it WFIRST} on a star-by-star bases, although there are only half as
many stars (due to half as much viewing area).  This is somewhat surprising
because of {\it Euclid}'s 4 times smaller aperture, 2.7 times lower throughput,
and 4 times fewer epochs.  However, these factors are partially compensated
by having 5 times longer exposures.  In addition, having a larger PSF
is actually helpful, since fewer photons wind up in saturated pixels.
In the end, however, the {\it Euclid} limit is about 0.8 mag brighter than
for {\it WFIRST}, but this still implies that it will obtain excellent
asteroseismology on about 100,000 stars.

We find that it is {\it Euclid}'s optical images that provide the best information
about KBOs.  These observations are auxiliary for the primary microlensing
program because there are fewer of them and the stars are significantly
extincted in the $I$ band.  The main microlensing use is to provide
colors, which are important in the interpretation of the events.  However,
they are especially useful for KBO observations, first because KBOs
(unlike stars) are not extincted, and second because the smaller PSF
(and pixels) reduces the background and improves the astrometry.

We find that {\it Euclid} will detect about 400 KBOs below the break and
about 530 per magnitude above the break, down to about $I=27.9$--28.4
(depending on whether the computational challenges can be solved).
At the break, the periods (and other orbital parameters) will be
measured to $\sim 2.5\%$.  At other magnitudes, the errors scale
inversely with flux.  {\it Euclid} will be roughly equally sensitive
(compared to {\it WFIRST}) to resolved binary companions, going down to
about $I=29.3$.  However, in contrast to {\it WFIRST}, it will not detect
many unresolved binaries via center-of-light motion.

\acknowledgments

Work by AG was supported by NSF grant AST 1103471 and
NASA grant NNX12AB99G.
DH acknowledges support by the Australian Research 
Council's Discovery Projects funding scheme (project number DE140101364) 
and support by NASA under Grant NNX14AB92G issued through the
Kepler Participating Scientist Program.
We thank Matthew Penny for seminal discussions.  
This research was greatly facilitated by the interactive environment at
the Galactic Archaeology Workshop at the Kavli Institute
for Theoretical Physics in Santa Barbara.

\end{document}